\documentclass[12pt,preprint]{aastex}

\providecommand{\jg}{J. Geophys.}
\providecommand{\pop}{Phys. Plas.}
\providecommand{\aipcp}{AIP Conf. Proc.}

\begin{document}
\title{Parallel Electric Field Spectrum of Solar Wind Turbulence}
\author{F.~S.~Mozer, C.~H.~K.~Chen}
\affil{Space Sciences Laboratory, University of California, Berkeley, California 94720, USA; fmozer@ssl.berkeley.edu}
\begin{abstract}
By searching through more than 10 satellite-years of \emph{THEMIS} and \emph{Cluster} data, three reliable examples of parallel electric field turbulence in the undisturbed solar wind have been found.  The perpendicular and parallel electric field spectra in these examples have similar shapes and amplitudes, even at large scales (frequencies below the ion gyroscale) where Alfv\'enic turbulence with no parallel electric field component is thought to dominate.  The spectra of the parallel electric field fluctuations are power laws with exponents near --5/3 below the ion scales ($\sim$0.1 Hz), and with a flattening of the spectrum in the vicinity of this frequency.  At small scales (above a few Hz), the spectra are steeper than --5/3 with values in the range of --2.1 to --2.8. These steeper slopes are consistent with expectations for kinetic Alfv\'en turbulence, although their amplitude relative to the perpendicular fluctuations is larger than expected.
\end{abstract}

\section{Introduction}

The magnetic field, velocity and density fluctuations in the solar wind have been measured for many decades \citep{coleman68,intriligator70} and the majority of our knowledge about solar wind turbulence has come from measurements of these quantities. Electric field fluctuations have only been used relatively recently \citep{bale05} but can provide important additional information about its nature, such as the phase speed of the fluctuations and the their possible dispersion properties. In this Letter, we present the first measurement of the electric field fluctuations of solar wind turbulence in the direction parallel to the magnetic field, and discuss the implications of the measurements.

At spacecraft-measured frequencies smaller than around 0.5 Hz, the power spectra of magnetic field, velocity and density fluctuations have a power law form with an exponent around --1.5 to --1.7 \citep{matthaeus82a,marsch90b,smith06a,chen11b,boldyrev11}. This is usually interpreted as inertial range turbulence, in which there is a cascade of energy from large to small scales. At higher frequencies, corresponding to convected scales smaller than the ion gyroscale and inertial length, the magnetic field and density spectra are seen to steepen \citep{denskat83,leamon98a,chen12a}, which is often attributed to damping \citep{goldstein94} or a further cascade of dispersive fluctuations \citep{ghosh96,schekochihin09}.

The first measurement of the electric field spectrum in the spacecraft frame shows that in the inertial range, it has a --5/3 scaling, the same as the magnetic field, and at high frequencies, around the ion gyroscale, may become enhanced compared to the magnetic field \citep{bale05}. The ratio of electric to magnetic field fluctuations was used to infer the phase speed, which was found to be consistent with kinetic Alfv\'en waves rather than whistler waves \citep{bale05,salem12}. Similar measurements have been made in the magnetosheath \citep{sundkvist07}, aurora \citep{chaston08} and magnetotail \citep{eastwood09}. Simulations of both gyrokinetic turbulence \citep{howes08b,howes11a} and Hall MHD turbulence \citep{dmitruk06,matthaeus08b,matthaeus10a} also show an enhancement in electric field fluctuations at ion kinetic scales, although comparison to the solar wind observations must be made carefully since, unlike the observations, the simulations are in a zero mean velocity frame.

It has been noted that it is important to consider the measurement frame when interpreting electric field measurements \citep{kellogg06}, since the electric field is not Galilean invariant. It has been shown that in the inertial range, the electric field in the spacecraft frame follows the same scaling as the magnetic field due to the fast convection past the spacecraft of the magnetic fluctuations \citep{chen11b}. In the frame of zero mean velocity, which is perhaps more interesting from a theoretical viewpoint, the electric field takes a shallower scaling of --1.4. 

For scales smaller than the ion kinetic scales, the perpendicular electric field spectrum in the zero mean velocity frame is predicted to have a scaling of $k_\perp^{-1/3}$, where $k_\perp$ is the wavevector perpendicular to the magnetic field direction \citep{schekochihin09}. A more recent model, which includes strong intermittency, predicts a scaling of $k_\perp^{-2/3}$ \citep{boldyrev12b}.

At scales larger than the ion kinetic scales, the turbulence is thought to be predominantly Alfv\'enic \citep{belcher71,bale05}, with no electric field fluctuations parallel to the magnetic field, although such fluctuations may arise from the compressive component of the turbulence. At smaller scales, i.e. frequencies greater than 0.5 Hz the Alfv\'enic turbulence is thought to develop parallel electric field fluctuations \citep{bian10a,bian10b}. Understanding this parallel electric field is important for several reasons. At the ion and electron kinetic scales it is thought that the turbulence is damped, which may occur partly by Landau damping, in which the particles interact with the parallel electric field fluctuations \citep{quataert98}. These fluctuations may also be important for the non-thermal acceleration of particles \citep[][and references therein]{bian10b}. Finally, measuring these fluctuations gives us a new way to probe the fundamental physics of kinetic scale plasma turbulence.

\section{Data}

Criteria for a successful parallel electric field spectral measurement in the quiet solar wind are stringent.  A first requirement is that data be transmitted at a high data rate (8192 samples/second for \emph{THEMIS}) and high gain in order to obtain spectral information at frequencies above a few Hz.  This high data rate transmission occurred less than 0.1\% of the time.  A second criterion is that, during such high rate transmissions, the spacecraft cannot be in the tail, the foreshock, or the lunar wake.  For \emph{THEMIS-C} in lunar orbit, this happened about 20\% of the time.  A next requirement is that the magnetic field vector must be essentially in the spacecraft spin plane (which is approximately the ecliptic plane for \emph{THEMIS} and \emph{Cluster}).  This is because electric field measurements along the spin axis were not made on \emph{Cluster} and were not of sufficient quality on \emph{THEMIS} due to its short on-axis antennas.  The magnetic field was in the spin plane infrequently, perhaps a few percent of the time.  A last requirement is that the spin plane magnetic field be perpendicular to the Sun-Earth line (in approximately the ecliptic Y-direction) because the sunward electric field component is perturbed by each of the four sensors rotating into the wake of the solar wind flow every spin period.  The combination of all these requirements led to finding two \emph{THEMIS} examples and one \emph{Cluster} example in searching about three years of data from the four \emph{Cluster} and five \emph{THEMIS} spacecraft.  This data is presented below.

Parallel electric field spectra were obtained in two different ways with the requirement of a good measurement being that the two spectral estimates agreed.  These two estimates were made in two different coordinate systems, the despun spacecraft coordinates (dsc) and the field-aligned coordinates (fac). The dsc coordinates are aligned with spacecraft axes such that Zdsc is along the spin axis (approximately the ecliptic normal), Xdsc is in the spin plane and pointing sunward (approximately in the X-direction in GSE coordinates), and Ydsc, also in the spin plane, pointing generally westward.  Because the Ydsc direction is the direction of the magnetic field in the three successful events, the parallel electric field in these events was also in the Ydsc direction. By contrast, in the field- aligned coordinate system, Zfac is the direction parallel to the instantaneous magnetic field, Xfac is perpendicular to the instantaneous magnetic field, pointing generally sunward, and Yfac is the third direction in a right-handed coordinate system. Thus, on average, when the magnetic field is nearly in the Ydsc direction, EYdsc should be equal to EZfac, but they may be unequal in detail for the following reasons. The field-aligned coordinate system fluctuates in space because the magnetic field direction varies. These angular fluctuations were less than one or a few degrees for the examples shown. On the other hand, EYdsc is in a coordinate system fixed in space.  Even so, EYdsc may contain components of the perpendicular electric field because the average and instantaneous directions of the magnetic field can be different. That these differences between EYdsc and EZfac are small and that either measurement gives the parallel electric field spectrum will be shown in the figures that follow. It is noted that a Galilean transformation to the plasma frame has not been made because the parallel electric field is Galilean invariant.

Fig. 1a shows an example of magnetic field measurements \citep{auster08} on the \emph{THEMIS-C} spacecraft from 08:55:10 UT to 09:04:10 UT on April 21, 2012, when the spacecraft was approximately on the Sun-Earth line and 60 Earth radii upstream and the plasma beta was 0.54.  Because the magnetic field was in the Ydsc direction, this event satisfies the earlier selection criteria and the parallel electric field was in the Ydsc direction. Any leakage of the perpendicular electric field into the Ydsc component can be at most a few percent because the ratio of BXdsc or BZdsc to BYdsc in Fig. 1a is a few percent. (We shall see below that the parallel and perpendicular spectra are about equal so the parallel electric field spectrum cannot be due to leakage of the perpendicular electric field).  Fig. 1b shows EYdsc time-domain data \citep{bonnell08} at a rate of 128 samples/second during the ten minute interval.  It is noted that the turbulence level during the first part of the data interval was larger than near the end of the interval. This is not unusual for such time domain data. The spectral slopes through the first and last halves of the interval were the same, so the data presented below is the average over the entire interval.

Parallel (EYdsc and EZfac) and perpendicular (EXdsc) electric field spectra are shown in Fig. 2a for the time interval of Fig. 1. An important demonstration of the validity of these spectral measurements is that the two parallel spectra agree.  It is also noted that the perpendicular power spectrum, which is plotted a factor of 100 lower on the vertical scale to separate it from the roughly equal power parallel spectrum, had a magnitude and shape similar to that of the parallel spectrum.  Wake effects or other perturbations of the waveform of Fig. 1b were less than about 3 millivolts/meter (mV/m) and they were removed by notch filtering the waveform of Fig. 1b at the fundamental and 11 harmonics to remove 1.2\% of the frequency space between $10^{-5}$ and 64 Hz. Even so, it is possible that the spectra outside the region of these narrow spikes were perturbed by the presence of the wake. To check this possibility, half of the total data in the time domain waveform was removed by deleting regions around each spike, and the spectral shape was the same as when all the data was used.
  
Within the 10 minute time interval whose spectra cover below 0.01 to 5 Hz, data was also fed to the ground through a higher gain, higher data rate, AC coupled channel that produced the spectra at frequencies of 5 to 1000 Hz in Fig. 2a.  The spectra of the total magnetic field and the trace of its correlation spectral matrix are given in Fig. 2b, with fluxgate \citep{auster08} and search coil \citep{roux08} instruments providing, respectively, the lower and higher frequency portions of the data.  The peak in the magnetic field spectra near 50 Hz is due to a physical process whose discussion is beyond the scope of this paper.  This peak is also seen at times in the higher frequency electric field spectra and a time interval was chosen for analysis that did not contain this spectral peak.  Because the higher frequency data collection was triggered by an on-board algorithm that selected the largest amplitude events, the higher frequency spectra contained more power than their lower frequency counterparts.  To compensate for this effect, the powers in these higher frequency components were decreased in order that the high and low frequency data fit together.   It is important to note that the power in the perpendicular electric field is about equal to that in the parallel electric field, which shows that the observed parallel electric field does not result from contamination of the perpendicular electric field.

The power law exponents of different segments of the electric and magnetic field spectra are given near each segment of each plot in Fig. 2.  With the exception of the lowest frequency portion of the perpendicular electric field spectrum (which may differ from the other spectra due to inadequate data coverage), the low frequency spectra all have power law slopes of about --5/3.  This includes the parallel electric field spectra that, for pure Alfv\'enic turbulence, would have no power at these lowest frequencies.  All the spectra are relatively flat near the proton gyrofrequency (the vertical dotted curve) and have spectral slopes of about --5/3 above this frequency until, above a few Hz, the spectra become steeper and have power law slopes between about --1.9 and --2.8.  The three vertical dotted lines are, from lowest to highest frequency, the proton gyrofrequency, the proton gyroscale (the solar wind velocity divided by $2\pi$ times the proton gyroradius), and the electron gyroscale. 

A second example of electric and magnetic field spectra is given in Fig. 3 for \emph{THEMIS-C} data collected on October 14, 2012 from 10:27 UT to 10:38 UT when the plasma beta was 1.1.  The features of these plots are the same as those described for the data of Fig. 2.  Namely, the two estimates of the parallel electric field power agree, the parallel spectrum has somewhat more power than the perpendicular spectrum, and the three power spectra have the same shape.  It is emphasized again that the parallel electric field power at the lowest frequencies was very much non-zero.  Also, the parallel electric field, perpendicular electric field, magnetic field trace, and total magnetic field all have power law slopes with spectral index about --5/3 at the lowest frequencies.  The spectra are all flatter near the ion gyroscale and have slopes of about --5/3 above this frequency until, above a frequency of a few Hz, the slopes become steeper to lie between about --2.1 and --2.5.   

Spectral data from \emph{Cluster-3} electric \citep{gustafsson97} and magnetic field \citep{balogh97} measurements covering 11.5 minutes at a distance of 19 Earth radii, a magnetic latitude of 19 degrees, and a magnetic local time of 10:15, are given in Fig. 3. For this case also, the magnetic field was in the Ydsc direction and the two parallel electric field spectra in Fig. 3a are in essential agreement.  Because the antennas on \emph{Cluster} were 88 meters, tip-to-tip, wake perturbations were much smaller than those observed on \emph{THEMIS} whose pair of spin plane antennas were 40 and 50 meters, tip-to-tip. Nevertheless, precautions taken in treating wake effects in the \emph{THEMIS} data were also used in analyzing the \emph{Cluster} event.  \emph{Cluster} was in the undisturbed solar wind during this time interval, as expected from the magnetic field being in the Y-direction and as shown by the lack of foreshock plasma in the electrostatic analyzers and the presence of Langmuir turbulence in the Whisper instrument.  The frequency coverage for this event extended to only 1 Hz because there was no higher frequency data collection.  For this event, the parallel electric field power was an order-of-magnitude larger than the perpendicular electric field power.  Otherwise the parallel and perpendicular spectra were essentially identical and they had power law slopes over the full frequency range of about --5/3 except in the vicinity of the ion gyrofrequency where the slopes were flatter.

\section{Discussion}

For the three examples of parallel electric field measurements in the solar wind, the parallel electric field estimated as EYdsc agreed with that estimated as EZfac, thereby validating the methods used to obtain the parallel electric field turbulence spectra. In general, the parallel electric field powers were comparable to or greater than the perpendicular powers, which rules out the possibility that the parallel fields were due to contamination from the perpendicular fields.  Both the electric and magnetic field spectra exhibited plateaus near the ion gyrofrequency ($\sim$0.1 Hz) and had exponential slopes of about --5/3 below and above this frequency.  The two events having spectra that extended to higher frequencies had spectra above a few Hz that became steeper with slopes between --1.9 and --2.8.

For low frequencies, below 0.1 Hz, the slope of the parallel electric field spectrum matches that of the magnetic field, which is not expected for pure Alfv\'enic turbulence \citep{bian10a,bian10b}. This may be due to the parallel electric field fluctuations from the slow mode like component of the turbulence, which contains around 10\% of the energy \citep[][and references therein]{howes12a} or could be due to the non-linear turbulence differing from the predictions of the linear wave modes. Similarly to the density spectrum \citep{chen12a,chen13a}, the parallel electric field spectrum also displays a local flattening around the ion scales. It is possible that this is for a similar reason to that discussed by \citet{chen13a}: the parallel electric field fluctuations from the kinetic Alfv\'en turbulence take over from the slow mode at this scale, causing the observed plateau. Smaller plateaus are also seen in the trace magnetic field spectra; the reason for this is not known, since they are not usually seen, although this feature may be related to the short interval length, meaning that the low frequencies are not well measured.

The parallel electric field spectrum at high frequencies is steeper than some predictions \citep{bian10a,bian10b} although it is in a similar range to some recent simulations of strong kinetic Alfv\'en turbulence (S. Boldyrev, private communication). Its amplitude relative to the perpendicular spectrum, however, is larger than expected. This may be due to other types of fluctuations being present at these scales, or to the non-linear turbulence differing from the linear wave mode predictions. The presence of the significant parallel electric field fluctuations has implications for how the turbulence heats the solar wind; in particular, it means that in addition to transit-time damping from the parallel magnetic field fluctuations, there may be significant Landau damping \citep{quataert98}.

In the future it would be desirable to obtain a larger number of intervals to check whether the results presented here are valid at a statistical level. Due to the lack of current data that meets the stringent conditions required for the analysis, a dedicated solar wind turbulence mission may be required for this.

\acknowledgments
This work was supported by NASA contracts NAS5-02099-09/12, NNN06AA01C and NASA grants NNX13AE24G, NNX09AE41G-1/14.

\begin{figure}
\begin{center}
\includegraphics[scale=0.9]{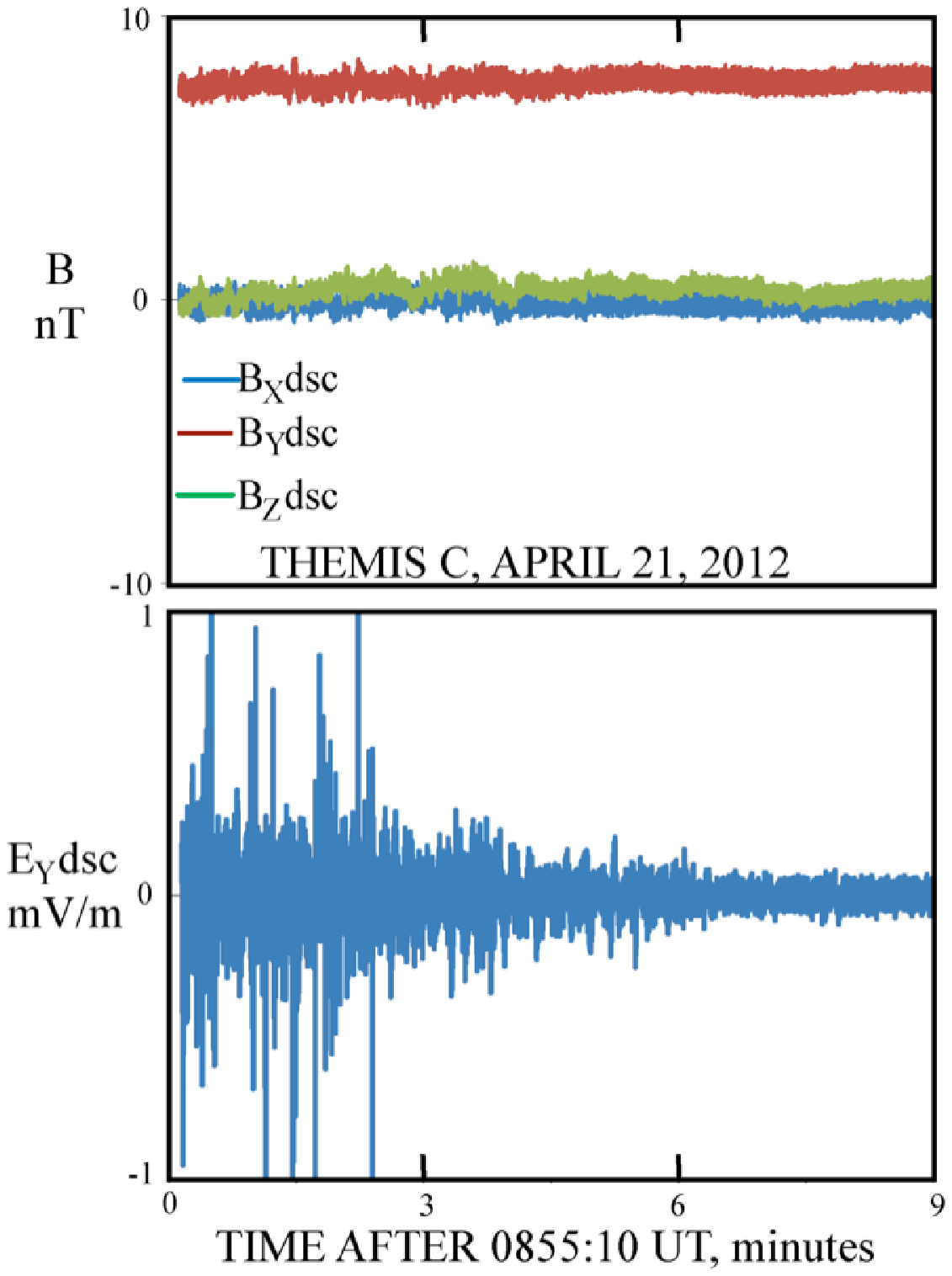}
\caption{\label{fig:f1}Three components of the magnetic field (Fig. 1a) and the Y-component of the electric field (Fig. 1b) during \emph{THEMIS-C} spacecraft observations from 08:55:10 UT to 09:04:10 UT on April 21, 2012, when the spacecraft was approximately on the Sun-Earth line and 60 Earth radii upstream. Because the magnetic field was in the Y-direction, the Y-component of the electric field was the parallel component of the electric field.}
\end{center}
\end{figure}

\begin{figure}
\begin{center}
\includegraphics[scale=0.9]{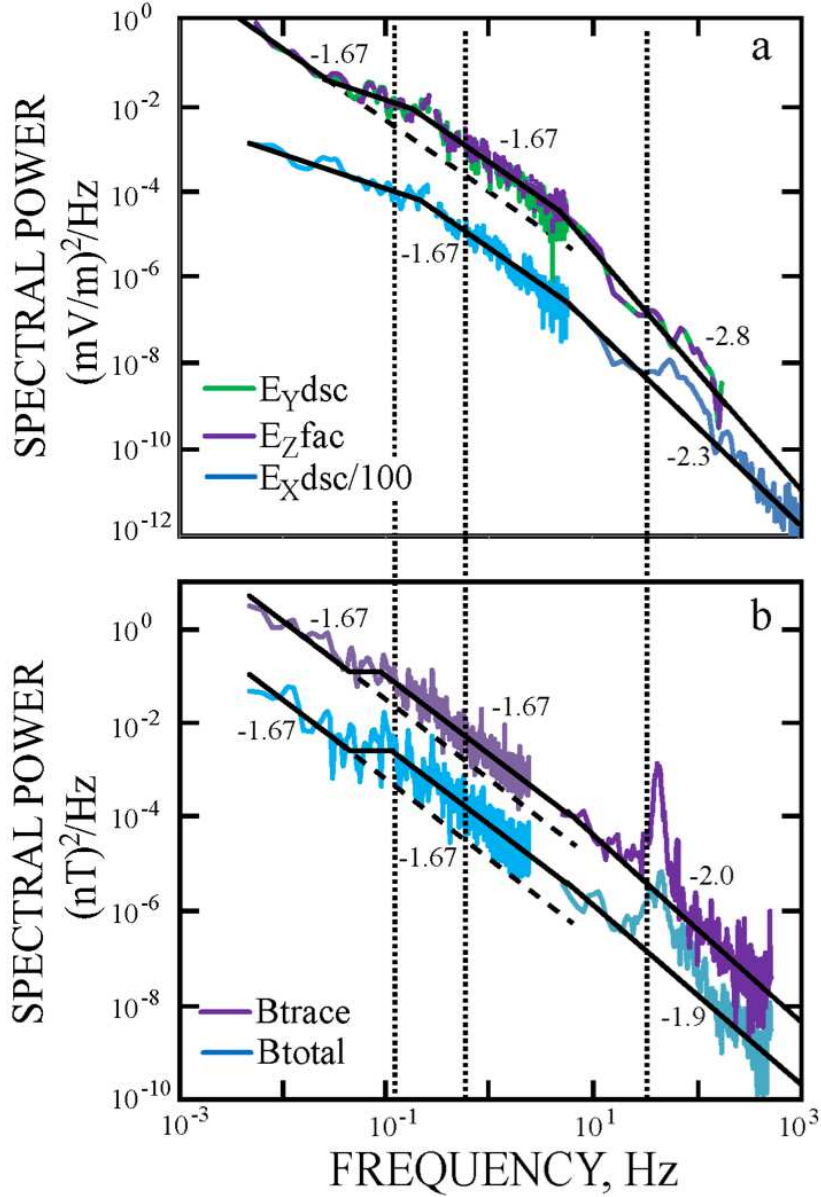}
\caption{\label{fig:f2}Spectra of the parallel electric field measured two different ways, the perpendicular electric field, the magnetic field trace and the total magnetic field during the data period illustrated in Fig. 1. The validity of the measurement is shown by the fact that the two parallel electric field spectra agree. In addition, the parallel and perpendicular electric field spectra have the same magnitudes, showing that the parallel electric field did not arise from contamination from the perpendicular electric field.  The power law slopes (the numbers near each segment of the plots) are in agreement with expectations except that the parallel electric field spectral amplitude is unexpectedly large. The three vertical dotted lines are, from lowest to highest frequency, the proton gyrofrequency, the proton gyroscale, and the electron gyroscale.}
\end{center}
\end{figure}

\begin{figure}
\begin{center}
\includegraphics[scale=0.9]{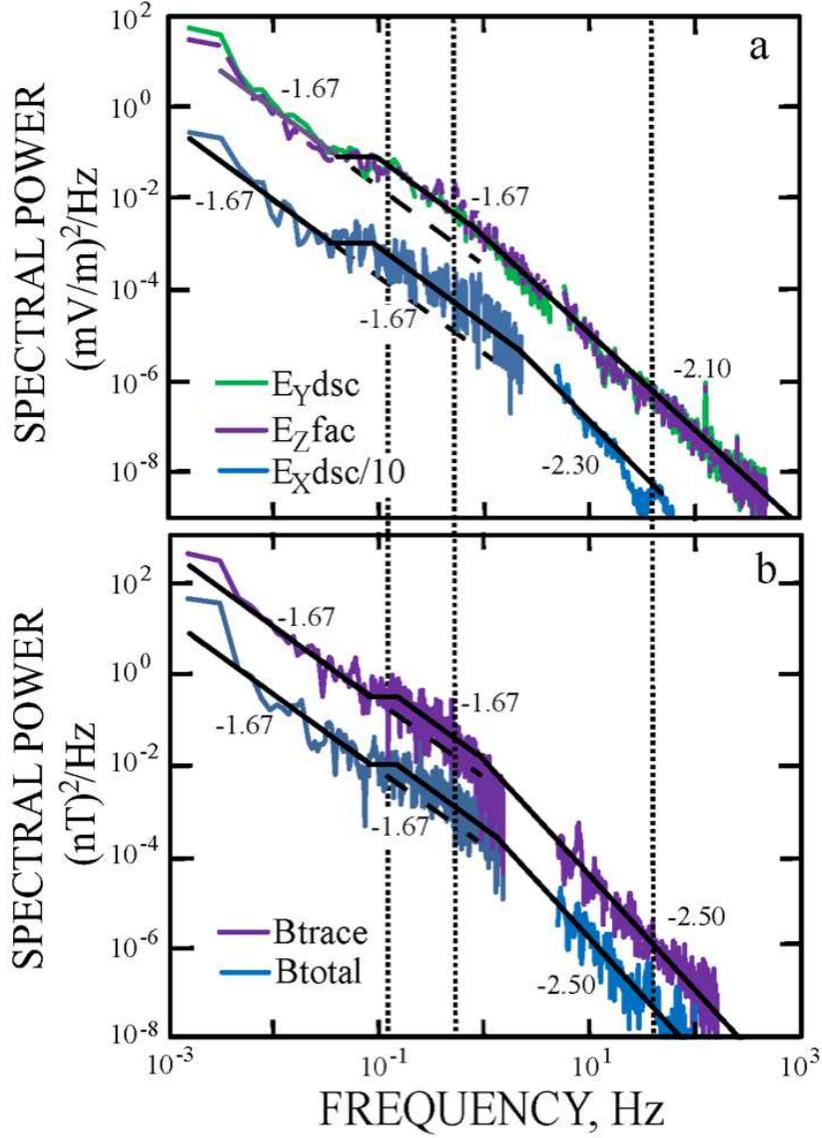}
\caption{\label{fig:f3}Spectra similar to those in Figure 2 for data collected on \emph{THEMIS-C} in the undisturbed solar wind on October 14, 2012 from 10:27 to 10:38 UT. The conclusions from Figure 2 are the same as those found from this figure.}
\end{center}
\end{figure}

\begin{figure}
\begin{center}
\includegraphics[scale=0.9]{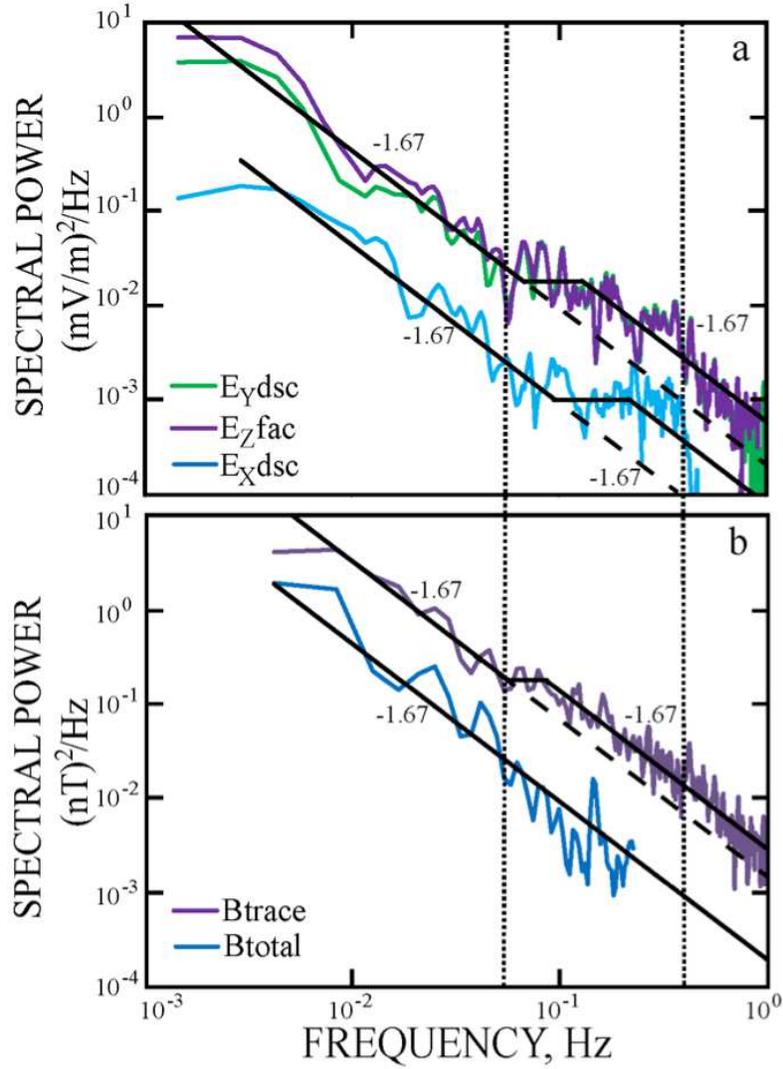}
\caption{\label{fig:f4}Spectra similar to those in Figures 2 and 3 for data collected in the undisturbed solar wind by the \emph{Cluster-3} satellite on February 10, 2001 06:17:30 to 06:29 UT. The conclusions from Figures 2 and 3 are the same as those found from this figure.}
\end{center}
\end{figure}


\begin{thebibliography}{36}
\expandafter\ifx\csname natexlab\endcsname\relax\def\natexlab#1{#1}\fi

\bibitem[{{Auster} {et~al.}(2008){Auster}, {Glassmeier}, {Magnes}, {Aydogar},
  {Baumjohann}, {Constantinescu}, {Fischer}, {Fornacon}, {Georgescu}, {Harvey},
  {Hillenmaier}, {Kroth}, {Ludlam}, {Narita}, {Nakamura}, {Okrafka},
  {Plaschke}, {Richter}, {Schwarzl}, {Stoll}, {Valavanoglou}, \&
  {Wiedemann}}]{auster08}
{Auster}, H.~U., {Glassmeier}, K.~H., {Magnes}, W., {et~al.} 2008, \ssr, 141,
  235

\bibitem[{{Bale} {et~al.}(2005){Bale}, {Kellogg}, {Mozer}, {Horbury}, \&
  {Reme}}]{bale05}
{Bale}, S.~D., {Kellogg}, P.~J., {Mozer}, F.~S., {Horbury}, T.~S., \& {Reme},
  H. 2005, \prl, 94, 215002

\bibitem[{{Balogh} {et~al.}(1997){Balogh}, {Dunlop}, {Cowley}, {Southwood},
  {Thomlinson}, {Glassmeier}, {Musmann}, {Luhr}, {Buchert}, {Acuna},
  {Fairfield}, {Slavin}, {Riedler}, {Schwingenschuh}, \& {Kivelson}}]{balogh97}
{Balogh}, A., {Dunlop}, M.~W., {Cowley}, S.~W.~H., {et~al.} 1997, \ssr, 79, 65

\bibitem[{{Belcher} \& {Davis}(1971)}]{belcher71}
{Belcher}, J.~W., \& {Davis}, L. 1971, \jgr, 76, 3534

\bibitem[{{Bian} \& {Kontar}(2010)}]{bian10a}
{Bian}, N.~H., \& {Kontar}, E.~P. 2010, \pop, 17, 062308

\bibitem[{{Bian} {et~al.}(2010){Bian}, {Kontar}, \& {Brown}}]{bian10b}
{Bian}, N.~H., {Kontar}, E.~P., \& {Brown}, J.~C. 2010, \aap, 519, A114

\bibitem[{{Boldyrev} \& {Perez}(2012)}]{boldyrev12b}
{Boldyrev}, S., \& {Perez}, J.~C. 2012, \apjl, 758, L44

\bibitem[{{Boldyrev} {et~al.}(2011){Boldyrev}, {Perez}, {Borovsky}, \&
  {Podesta}}]{boldyrev11}
{Boldyrev}, S., {Perez}, J.~C., {Borovsky}, J.~E., \& {Podesta}, J.~J. 2011,
  \apjl, 741, L19

\bibitem[{{Bonnell} {et~al.}(2008){Bonnell}, {Mozer}, {Delory}, {Hull},
  {Ergun}, {Cully}, {Angelopoulos}, \& {Harvey}}]{bonnell08}
{Bonnell}, J.~W., {Mozer}, F.~S., {Delory}, G.~T., {et~al.} 2008, \ssr, 141,
  303

\bibitem[{{Chaston} {et~al.}(2008){Chaston}, {Salem}, {Bonnell}, {Carlson},
  {Ergun}, {Strangeway}, \& {McFadden}}]{chaston08}
{Chaston}, C.~C., {Salem}, C., {Bonnell}, J.~W., {et~al.} 2008, \prl, 100,
  175003

\bibitem[{{Chen} {et~al.}(2011){Chen}, {Bale}, {Salem}, \& {Mozer}}]{chen11b}
{Chen}, C.~H.~K., {Bale}, S.~D., {Salem}, C., \& {Mozer}, F.~S. 2011, \apjl,
  737, L41

\bibitem[{{Chen} {et~al.}(2013){Chen}, {Howes}, {Bonnell}, {Mozer}, {Klein}, \&
  {Bale}}]{chen13a}
{Chen}, C.~H.~K., {Howes}, G.~G., {Bonnell}, J.~W., {et~al.} 2013, Solar Wind
  13 Proc. (in press), arXiv:1210.0127

\bibitem[{{Chen} {et~al.}(2012){Chen}, {Salem}, {Bonnell}, {Mozer}, \&
  {Bale}}]{chen12a}
{Chen}, C.~H.~K., {Salem}, C.~S., {Bonnell}, J.~W., {Mozer}, F.~S., \& {Bale},
  S.~D. 2012, \prl, 109, 035001

\bibitem[{{Coleman}(1968)}]{coleman68}
{Coleman}, P.~J. 1968, \apj, 153, 371

\bibitem[{{Denskat} {et~al.}(1983){Denskat}, {Beinroth}, \&
  {Neubauer}}]{denskat83}
{Denskat}, K.~U., {Beinroth}, H.~J., \& {Neubauer}, F.~M. 1983, \jg, 54, 60

\bibitem[{{Dmitruk} \& {Matthaeus}(2006)}]{dmitruk06}
{Dmitruk}, P., \& {Matthaeus}, W.~H. 2006, \pop, 13, 042307

\bibitem[{{Eastwood} {et~al.}(2009){Eastwood}, {Phan}, {Bale}, \&
  {Tjulin}}]{eastwood09}
{Eastwood}, J.~P., {Phan}, T.~D., {Bale}, S.~D., \& {Tjulin}, A. 2009, \prl,
  102, 035001

\bibitem[{{Ghosh} {et~al.}(1996){Ghosh}, {Siregar}, {Roberts}, \&
  {Goldstein}}]{ghosh96}
{Ghosh}, S., {Siregar}, E., {Roberts}, D.~A., \& {Goldstein}, M.~L. 1996, \jgr,
  101, 2493

\bibitem[{{Goldstein} {et~al.}(1994){Goldstein}, {Roberts}, \&
  {Fitch}}]{goldstein94}
{Goldstein}, M.~L., {Roberts}, D.~A., \& {Fitch}, C.~A. 1994, \jgr, 99, 11519

\bibitem[{{Gustafsson} {et~al.}(1997){Gustafsson}, {Bostrom}, {Holback},
  {Holmgren}, {Lundgren}, {Stasiewicz}, {Ahlen}, {Mozer}, {Pankow}, {Harvey},
  {Berg}, {Ulrich}, {Pedersen}, {Schmidt}, {Butler}, {Fransen}, {Klinge},
  {Thomsen}, {Falthammar}, {Lindqvist}, {Christenson}, {Holtet}, {Lybekk},
  {Sten}, {Tanskanen}, {Lappalainen}, \& {Wygant}}]{gustafsson97}
{Gustafsson}, G., {Bostrom}, R., {Holback}, B., {et~al.} 1997, \ssr, 79, 137

\bibitem[{{Howes} {et~al.}(2012){Howes}, {Bale}, {Klein}, {Chen}, {Salem}, \&
  {TenBarge}}]{howes12a}
{Howes}, G.~G., {Bale}, S.~D., {Klein}, K.~G., {et~al.} 2012, \apjl, 753, L19

\bibitem[{{Howes} {et~al.}(2008){Howes}, {Dorland}, {Cowley}, {Hammett},
  {Quataert}, {Schekochihin}, \& {Tatsuno}}]{howes08b}
{Howes}, G.~G., {Dorland}, W., {Cowley}, S.~C., {et~al.} 2008, \prl, 100,
  065004

\bibitem[{{Howes} {et~al.}(2011){Howes}, {TenBarge}, {Dorland}, {Quataert},
  {Schekochihin}, {Numata}, \& {Tatsuno}}]{howes11a}
{Howes}, G.~G., {TenBarge}, J.~M., {Dorland}, W., {et~al.} 2011, \prl, 107,
  035004

\bibitem[{{Intriligator} \& {Wolfe}(1970)}]{intriligator70}
{Intriligator}, D.~S., \& {Wolfe}, J.~H. 1970, \apj, 162, L187

\bibitem[{{Kellogg} {et~al.}(2006){Kellogg}, {Bale}, {Mozer}, {Horbury}, \&
  {Reme}}]{kellogg06}
{Kellogg}, P.~J., {Bale}, S.~D., {Mozer}, F.~S., {Horbury}, T.~S., \& {Reme},
  H. 2006, \apj, 645, 704

\bibitem[{{Leamon} {et~al.}(1998){Leamon}, {Smith}, {Ness}, {Matthaeus}, \&
  {Wong}}]{leamon98a}
{Leamon}, R.~J., {Smith}, C.~W., {Ness}, N.~F., {Matthaeus}, W.~H., \& {Wong},
  H.~K. 1998, \jgr, 103, 4775

\bibitem[{{Marsch} \& {Tu}(1990)}]{marsch90b}
{Marsch}, E., \& {Tu}, C.-Y. 1990, \jgr, 95, 11945

\bibitem[{{Matthaeus} \& {Goldstein}(1982)}]{matthaeus82a}
{Matthaeus}, W.~H., \& {Goldstein}, M.~L. 1982, \jgr, 87, 6011

\bibitem[{{Matthaeus} {et~al.}(2008){Matthaeus}, {Servidio}, \&
  {Dmitruk}}]{matthaeus08b}
{Matthaeus}, W.~H., {Servidio}, S., \& {Dmitruk}, P. 2008, \prl, 101, 149501

\bibitem[{{Matthaeus} {et~al.}(2010){Matthaeus}, {Servidio}, \&
  {Dmitruk}}]{matthaeus10a}
{Matthaeus}, W.~H., {Servidio}, S., \& {Dmitruk}, P. 2010, \aipcp, 1216, 184

\bibitem[{{Quataert}(1998)}]{quataert98}
{Quataert}, E. 1998, \apj, 500, 978

\bibitem[{{Roux} {et~al.}(2008){Roux}, {Le Contel}, {Coillot}, {Bouabdellah},
  {de La Porte}, {Alison}, {Ruocco}, \& {Vassal}}]{roux08}
{Roux}, A., {Le Contel}, O., {Coillot}, C., {et~al.} 2008, \ssr, 141, 265

\bibitem[{{Salem} {et~al.}(2012){Salem}, {Howes}, {Sundkvist}, {Bale},
  {Chaston}, {Chen}, \& {Mozer}}]{salem12}
{Salem}, C.~S., {Howes}, G.~G., {Sundkvist}, D., {et~al.} 2012, \apjl, 745, L9

\bibitem[{{Schekochihin} {et~al.}(2009){Schekochihin}, {Cowley}, {Dorland},
  {Hammett}, {Howes}, {Quataert}, \& {Tatsuno}}]{schekochihin09}
{Schekochihin}, A.~A., {Cowley}, S.~C., {Dorland}, W., {et~al.} 2009, \apjs,
  182, 310

\bibitem[{{Smith} {et~al.}(2006){Smith}, {Hamilton}, {Vasquez}, \&
  {Leamon}}]{smith06a}
{Smith}, C.~W., {Hamilton}, K., {Vasquez}, B.~J., \& {Leamon}, R.~J. 2006,
  \apjl, 645, L85

\bibitem[{{Sundkvist} {et~al.}(2007){Sundkvist}, {Retin{\`o}}, {Vaivads}, \&
  {Bale}}]{sundkvist07}
{Sundkvist}, D., {Retin{\`o}}, A., {Vaivads}, A., \& {Bale}, S.~D. 2007, \prl,
  99, 025004

\end{thebibliography}
\end{document}